\pdfoutput=1
\documentclass[useAMS,usenatbib]{mn2e}
\usepackage{amssymb}
\usepackage{bbm}
\usepackage{txfonts}
\usepackage{psfrag}
\usepackage{graphicx}
\usepackage{natbib}
\usepackage{if then}
\usepackage{longtable}

\def\del#1{{}}

\newcommand{\unit}[1]{\mbox{  }\rm{#1}}

\title{Fast Hamiltonian Sampling for large scale structure inference}
\author[Jens Jasche \\$^{1}$, Francisco Shu Kitaura]
       {Jens Jasche $^{1}$, Francisco S. Kitaura $^{2}$ \\$^{1}$ Max-Planck-Institut f\"{u}r Astrophysik , Karl-Schwarzschild Stra\ss e 1,  D-85748 Garching, Germany\\$^{2}$ SISSA, Scuola Internazionale Superiore di Studi Avanzati , via Beirut 2-4,  I-34014 Trieste, Italy}
\begin{document}
\date{Submitted to MNRAS 21-Sept-2009}

\pagerange{\pageref{firstpage}--\pageref{lastpage}} \pubyear{2006}

\maketitle

\label{firstpage}

\begin{abstract}
In this work we present a new and efficient Bayesian method for nonlinear three dimensional large scale structure inference. We employ a Hamiltonian Monte Carlo (HMC) sampler to obtain samples from a multivariate highly non-Gaussian lognormal Poissonian density posterior given a set of observations. The HMC allows us to take into account the nonlinear relations between the observations and the underlying density field which we seek to recover. As the HMC provides a sampled representation of the density posterior any desired statistical summary, such as the mean, mode or variance, can be calculated from the set of samples. Further, it permits us to seamlessly propagate non-Gaussian uncertainty information to any final quantity inferred from the set of samples.
The developed method is extensively tested in a variety of test scenarios, taking into account a highly structured survey geometry and selection effects. Tests with a mock galaxy catalog based on the millennium run show that the method is able to recover the filamentary structure of the nonlinear density field. The results further demonstrate the feasibility of non-Gaussian sampling in high dimensional spaces, as required for precision nonlinear large scale structure inference.
The HMC is a flexible and efficient method, which permits for simple extension and incorporation of additional observational constraints. Thus, the method presented here provides an efficient and flexible basis for future high precision large scale structure inference.
\end{abstract}


\section{Introduction}
Modern large galaxy surveys allow us to probe cosmic large scale structure to very high accuracy if the enormous amount of data can be processed and analyzed efficiently. Especially, precision reconstruction of the three dimensional density field from observations poses complex numerical challenges.
For this reason, several reconstruction methods and attempts to recover the underlying density field from galaxy observations have been presented in literature \citep[see e.g.][]{1989ApJ...336L...5B,1991ASPC...15...67B,1994ApJ...423L..93L,HOFFMAN1994,1995MNRAS.272..885F,BISTOLAS1998,WEBSTER1997,1999AJ....118.1146S,2002MNRAS.331..901Z,2004MNRAS.352..939E,KITAURA2009,JASCHE2009}.
Recently, \citet{KITAURA2009} presented a high resolution Wiener reconstruction of the Sloan Digital Sky Survey (SDSS) matter density field, and demonstrated the feasibility of precision large scale structure analysis. The Wiener filtering approach is based on a linear data model, which takes into account several observational effects, such as survey geometry, selection effects and noise \citep{kitaura,KITAURA2009,JASCHE2009}.
Although, the Wiener filter has proven to be extremely efficient for three dimensional matter field reconstruction, it still relies on a Gaussian approximation of the density posterior.
While this is an adequate approximation for the largest scales, precision recovery of nonlinear density structures may require non-Gaussian posteriors.
Especially, the detailed treatment of the non-Gaussian behavior and structure of the Poissonian shot noise contribution may allow for more precise recovery of poorly sampled objects. In addition, for a long time it has been suggested that the fully evolved nonlinear matter field can be well described by lognormal statistics \citep[see e.g.][]{HUBBLE1934,PEEBLES1980,COLES1991,GAZTANAGA1993,KAYO2001}. These discussions seem to advocate the use of a lognormal Poissonian posterior for large scale structure inference.
Several methods have been proposed to take into account non-Gaussian density posteriors \citep[see e.g.][]{SAUNDERS2000,kitaura,ENSSLIN2008,KJ2009}.

However, if the recovered nonlinear density field is to be used for scientific purposes, the method not only has to provide a single estimate, such as a mean or maximum a postiori reconstruction, but it should also provide uncertainty information, and the means to nonlinearly propagate this uncertainty to any final quantity inferred from the recovered density field.

For this reason, here we propose a new Bayesian method for nonlinear large scale structure inference. The developed computer program HADES (HAmiltonian Density Estimation and Sampling) explores the posterior distribution via an Hamiltonian Monte Carlo (HMC) sampling scheme. Unlike conventional Metropolis Hastings algorithms, which move through the parameter space by a random walk, and therefore require prohibitive amounts of steps to explore high dimensional spaces, the HMC sampler suppresses random walk behavior by introducing a persistent motion of the Markov chain through the parameter space \citep[][]{DUANE1987,NEAL1993,NEAL1996}. In this fashion, the HMC sampler maintains a reasonable efficiency even for high dimensional problems \citep{HANSON2001}. The HMC sampler has been widely used in Bayesian computation \citep[see e.g.][]{NEAL1993}. In cosmology it has been employed for cosmological parameter estimation and CMB data analysis \citep{HAJIAN2007,TAYLOR2008}.

In this work we demonstrate that the HMC can efficiently be used to sample the lognormal Poissonian posterior even in high dimensional spaces. In this fashion, the method is able to take into account the nonlinear relationship between the observation and the underlying density which we seek to recover.
The scientific output of the HMC is a sampled representation of the density posterior. For this reason, any desired statistical summary such as mean, mode and variance can easily be calculated from the HMC samples. Further, the full non-Gaussian uncertainty information can seamlessly be propagated to any finally estimated quantity by simply applying the according estimation procedure to all samples. This allows us to estimate the accuracy of conclusions drawn from the analyzed data.

In this work, we begin, in section \ref{PDF_LOGNORMAL}, by presenting a short justification for the use of the lognormal distribution as a prior for nonlinear density inference, followed by a discussion of the lognormal Poissonian posterior in section \ref{PDF_LOGPOIS}. Section \ref{HAMILTONIAN_SAMPLING} outlines the HMC method. In section \ref{equations_of_motion} the Hamiltonian equations of motion for the lognormal Poissonian posterior are presented. Details of the numerical implementation are described in section \ref{Numerical_implementation}. 
The method is extensively tested in section \ref{HADES_testing} by applying HADES to generated mock observations, taking into account a highly structured survey geometry and selection effects.
In section \ref{Discussion} we summarize and conclude.

\section{The lognormal distribution of density}
\label{PDF_LOGNORMAL}
In standard cosmological pictures, it is assumed that the initial seed perturbations in the primordial density field originated from an inflationary phase in the early stages of the Big Bang. This inflationary phase enhances microscopic quantum fluctuations to macroscopic scales yielding the initial density fluctuations required for gravitational collapse. These theories predict the initial density field amplitudes to be Gaussian distributed.
However, it is obvious that Gaussianity of the density field can only be true in the limit \(|\delta|\ll 1\), where \(\delta\) is the density contrast. In fully evolved density fields with amplitudes of \(\sigma_8>1\), as observed in the sky at scales of galaxies, clusters and super clusters, a Gaussian density distribution would allow for negative densities, and therefore would violate weak and strong energy conditions. In particular, it would give rise to negative mass (\(\delta < -1\)).
Therefore, in the course of gravitational structure formation the density field must have changed its statistical properties.
\citet{COLES1991} argue that assuming Gaussian initial conditions in the density and velocity distributions will lead to a log-normally distributed density field. It is a direct consequence of the continuity equation or the conservation of mass. 

Although, the exact probability distribution for the density field in nonlinear regimes is not known, the lognormal distribution seems to be a good phenomenological guess with a long history. Already Hubble noticed that galaxy counts in two dimensional cells on the sky can be well approximated by a lognormal distribution \citep{HUBBLE1934}. Subsequently, the lognormal distribution has been extensively discussed and agreements with galaxy observations have been found \citep[e.g.][]{HUBBLE1934,PEEBLES1980,COLES1991,GAZTANAGA1993,KAYO2001}.
\citet{KAYO2001} studied the probability distribution of cosmological nonlinear density fluctuations from N-body simulations with Gaussian initial conditions. They found that the lognormal distribution accurately describes the nonlinear density field even up to values of the density contrast of \(\delta \sim 100\).

Therefore, according to observations and theoretical considerations, we believe, that the statistical behavior of the nonlinear density field can be well described by a multivariate lognormal distribution, as given by:
\begin{equation}
\label{eq:LogNormal_prior}
{\mathcal P}(\{s_k\}|Q)=\frac{1}{\sqrt{2\pi det(Q)}} e^{-\frac{1}{2}\sum_{ij} \left(ln(1+s_i)+\mu_i\right) Q^{-1}_{ij} \left(ln(1+s_j)+\mu_j\right)} \prod_k \frac{1}{1+s_k} \, ,
\end{equation}
where \(Q\) is the covariance matrix of the lognormal distribution and \(\mu_i\) describes a constant mean field given by:
\begin{equation}
\label{eq:MU}
\mu_i=\frac{1}{2}\sum_{i,j} Q_{ij} \, .
\end{equation}
This probability distribution, seems to be an adequate prior choice for reconstructing the present density field. However, using such a prior requires highly nonlinear reconstruction methods, as will be presented in the following.

\section{Lognormal Poissonian posterior}
\label{PDF_LOGPOIS}
Studying the actual matter distribution of the Universe requires to draw inference from some observable tracer particle. The most obvious tracer particles for the cosmic density field are galaxies, which tend to follow the gravitational potential of matter. As galaxies are discrete particles, the galaxy distribution can be described as a specific realization drawn from an inhomogeneous Poisson process \citep[see e.g.][]{LAYZER1956,PEEBLES1980,MARTINEZ2002}. The according probability distribution is given as:
\begin{equation}
\label{eq:Poissonian}
{\mathcal P}(\{N_k^{g}\}|\{\lambda_k\})= \prod_k \frac{{\left(\lambda_k\right)}^{N^{g}_k} e^{-\lambda_k}}{{N^{g}_k}!} \, ,
\end{equation}
where \(N_k^{g}\) is the observed galaxy number at position \(\vec{x}_k\) in the sky and  \(\lambda_k\) is the expected number of galaxies at this position.
The mean galaxy number is related to the signal \(s_k\) via:
\begin{equation}
\label{eq:data_model}
\lambda_k= R_k \bar{N}(1+B(s)_k)\, ,
\end{equation}
where \(R_k\) is a linear response operator, incorporating survey geometries and selection effects, \(\bar{N}\) is the mean number of galaxies in the volume and \(B(x)_k\) is a nonlinear, non local, bias operator at position \(\vec{x}_k\).
The lognormal prior given in equation (\ref{eq:LogNormal_prior}) together with the Poissonian likelihood given in equation (\ref{eq:Poissonian}) yields the lognormal Poissonian posterior, for the density contrast \(s_k\) given some galaxy observations \(N_k^{g}\):
\begin{eqnarray}
\label{eq:LOGNORMALPOISSONIAN_POSTERIOR}
{\mathcal P}(\{s_k\}|\{N_k^{g}\}) &=& \frac{e^{-\frac{1}{2}\sum_{ij} \left(ln(1+s_i)+\mu_i\right) Q^{-1}_{ij} \left(ln(1+s_j)+\mu_j\right)}}{\sqrt{2\pi det(Q)}} \prod_l \frac{1}{1+s_l} \nonumber \\
& & \times \prod_k \frac{{\left(R_k \bar{N}(1+B(s)_k)\right)}^{N^{g}_k} e^{-R_k \bar{N}(1+B(s)_k)}}{{N^{g}_k}!}
\end{eqnarray}
However, this posterior greatly simplifies if we perform the change of variables by introducing \(r_k=ln(1+s_k)\). Note, that this change of variables is also numerically advantageous, as it prevents numerical instabilities at values \(\delta \sim -1\). Hence, we yield the Posterior
\begin{eqnarray}
\label{eq:LOGNORMALPOISSONIAN_POSTERIOR_R}
{\mathcal P}(\{r_k\}|\{N_k^{g}\}) &=& \frac{e^{-\frac{1}{2}\sum_{ij} \left(r_i+\mu_i\right) Q^{-1}_{ij} \left(r_j+\mu_j\right)}}{\sqrt{2\pi det(Q)}} \nonumber \\ & & \times  \prod_k \frac{{\left(R_k \bar{N}(1+B(e^{r}-1)_k)\right)}^{N^{g}_k} e^{-R_k \bar{N}(1+B(e^{r}-1)_k)}}{{N^{g}_k}!} \, . \nonumber \\
\end{eqnarray}
It is important to note, that this is a highly non-Gaussian distribution, and nonlinear reconstruction methods are required in order to perform accurate matter field reconstructions in the nonlinear regime. In example, estimating the maximum a postiori values from the lognormal Poissonian distribution involves the solution of implicit equations. However, we are not solely interested in a single estimate of the density distribution, we rather prefer to draw samples from the lognormal Poissonian posterior. In the following, we are therefore describing a numerically efficient method to sample from this highly non-Gaussian distribution.

\section{Hamiltonian sampling}
\label{HAMILTONIAN_SAMPLING}
As already described in the previous section the lognormal Poissonian posterior will involve highly nonlinear reconstruction methods and will therefore be numerically demanding.
Nevertheless, since we propose a Bayesian method, we are not interested in only providing a single estimate of the density field, but would rather be able to sample from the full non-Gaussian posterior.
Unlike, in the Gibbs sampling approach to density field sampling, as proposed in \citet{JASCHE2009}, there unfortunately exists no known way to directly draw samples from the lognormal Poissonian distribution. For this reason, a Metropolis-Hastings sampling mechanism has to be employed.
 
However, the Metropolis-Hastings has the numerical disadvantage that not every sample will be accepted. A low acceptance rate can therefore result in a prohibitive numerical scaling for the method, especially since we are interested in estimating full three dimensional matter fields which usually have about \(10^6\) or more free parameters \(s_k\). This high rejection rate is due to the fact, that conventional Markov Chain Monte Carlo (MCMC) methods move through the parameter space by a random walk and therefore require a prohibitive amount of samples to explore high-dimensional spaces.
Given this situation, we propose to use a Hybrid Monte Carlo method, which in the absence of numerical errors, would yield an acceptance rate of unity.

The so called Hamiltonian Monte Carlo (HMC) method exploits techniques developed to follow classical dynamical particle motion in potentials \citep[][]{DUANE1987,NEAL1993,NEAL1996}. In this fashion the Markov sampler follows a persistent motion through the parameter space, supressing the random walk behavior. This enables us to sample with reasonable efficiency in high dimensional spaces \citep[][]{HANSON2001}.

The idea of the Hamiltonian sampling can be easily explained.
Suppose, that we wish to draw samples from the probability distribution \({\mathcal P}(\{x_i\})\), where \(\{x_i\}\) is a set consisting of the \(N\) elements \(x_i\). If we interpret the negative logarithm of this posterior distribution as a potential:
\begin{equation}
\label{eq:Potential}
\psi(x)=-ln({\mathcal P}(x)) \, ,
\end{equation}
and by introducing a 'momentum' variable \(p_i\) and a 'mass matrix' \(M\), as nuisance parameters, we can formulate a Hamiltonian describing the dynamics in the multi dimensional phase space.
Such a Hamiltonian is then given as:
\begin{equation}
\label{eq:Hamiltonian}
H = \sum_i\sum_j \frac{1}{2}\,p_i\,M_{ij}^{-1}\,p_j +\psi(x) \, ,
\end{equation}
As can be seen in equation (\ref{eq:Hamiltonian}), the form of the Hamiltonian is such, that this distribution is separable into a Gaussian distribution in the momenta \(\{p_i\}\) and the target distribution \({\mathcal P}(\{x_i\})\) as:
\begin{equation}
\label{eq:TARGET_DISTRIBUTION}
e^{-H} = {\mathcal P}(\{x_i\})\,e^{-\frac{1}{2}\,\sum_i\sum_j\,p_i\,M_{ij}^{-1}\,p_j}\, .
\end{equation}
It is therefore obvious that, marginalizing over all momenta will yield again our original target distribution \({\mathcal P}(\{x_i\})\).

Our task now is to draw samples from the joint distribution, which is proportional to \(exp(-H)\).
To find a new sample of the joint distribution we first draw a set of momenta from the distribution defined by the kinetic energy term, that is an \(N\) dimensional Gaussian with a covariance matrix \(M\). We then allow our system to evolve deterministically, from our starting point \( (\{x_i\},\{p_i\})\) in the phase space for some fixed pseudo time \(\tau\) according to Hamilton's equations:
\begin{equation}
\label{eq:HAMILTON1}
\frac{dx_i}{dt} = \frac{\partial H}{\partial p_i}\, .
\end{equation}

\begin{equation}
\label{eq:HAMILTON2}
\frac{dp_i}{dt} = \frac{\partial H}{\partial x_i} = - \frac{\partial \psi(x)}{\partial x_i}\, .
\end{equation}
The integration of this equations of motion yields the new position \((\{x'_i\},\{p'_i\})\) in phase space. This new point is accepted according to the usual acceptance rule:
\begin{equation}
\label{eq:acceptance_rule}
{\mathcal P}_A = min\left[1,exp(-\left(H(\{x'_i\},\{p'_i\})-H(\{x_i\},\{p_i\})\right)\right]\, .
\end{equation}
Since the equations of motion provide a solution to a Hamiltonian system, energy or the Hamiltonian given in equation (\ref{eq:Hamiltonian}) is conserved, and therefore the solution to this system provides an acceptance rate of unity. In practice however, numerical errors can lead to a somewhat lower acceptance rate.
Once a new sample has been accepted the momentum variable is discarded and the process restarts by randomly drawing a new set of momenta.
The individual momenta \(\{p_i\}\) will not be stored, and therefore discarding them amounts to marginalizing over this auxiliary quantity.
Hence, the Hamiltonian sampling procedure basically consists of two steps. The first step is a Gibbs sampling step, which yields a new set of Gaussian distributed momenta. The second step, on the other hand amounts to solving a dynamical trajectory on the posterior surface.
%
%
%
%

\begin{figure*}
\centering{\resizebox{1.\hsize}{!}{\includegraphics[width=0.5\textwidth,clip=true]{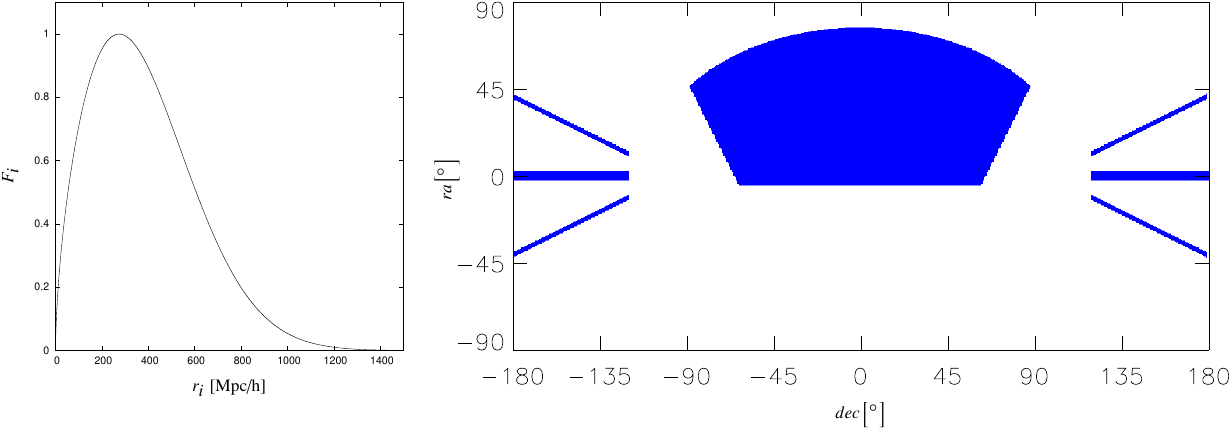}}}
\caption{Selection function and two dimensional sky mask used for the generation of mock galaxy observations.}
	\label{fig:TEST_SEL_WIN}
\end{figure*}

\section{Equations of motion for a log-normal Poissonian system}
\label{equations_of_motion}
In the framework of Hamiltonian sampling the task of sampling from the lognormal Poissonian posterior reduces to solving the corresponding Hamiltonian system.
Given the posterior distribution defined in equation (\ref{eq:LOGNORMALPOISSONIAN_POSTERIOR_R}) we can write the potential \(\psi(\{r_k\})\) as:
\begin{eqnarray}
\label{eq:LOGPOISSONIAN_POTENTIAL}
 \psi(\{r_k\}) &=& \frac{1}{2} ln(2\pi det(Q)) \nonumber \\
& + & \frac{1}{2}\sum_{ij} \left(r_i+\mu_i\right) Q^{-1}_{ij} \left(r_j+\mu_j\right) \nonumber \\
& - & \sum_k \left [ ln\left(\frac{(R_k\bar{N})^{N^{g}_k}}{{N^{g}_k}!}\right) + N^{g}_k\,ln(1+B(e^{r}-1)_k) \right .\nonumber \\
&  & \left . - R_k\bar{N}(1+B(e^{r}-1)_k) \right ]\, . \nonumber \\
\end{eqnarray}
The gradient of this potential with respect to \(r_l\) then yields the forces, given as:
\begin{eqnarray}
\label{eq:LOGPOISSONIAN_FORCES}
\frac{\partial \psi(\{r_k\})}{\partial r_l} &=& \sum_{j} Q^{-1}_{lj} \left(r_j+\mu_j\right) \nonumber \\
& & - \left( \frac{N^{g}_l}{(1+B(e^{r}-1)_l)} - R_l\bar{N}\right)\left .\frac{\partial B(e^{r}-1)}{\partial(e^{r}-1)}\right |_l \, e^{r_l} \, . \nonumber \\ 
\end{eqnarray}
Equation (\ref{eq:LOGPOISSONIAN_FORCES}) obviously is a very general formulation of the reconstruction problem, and it demonstrates that the Hamiltonian sampler can in principle deal with all kinds of nonlinearities, especially in the case of the bias operator \(B(x)\).
However, for the sake of this paper, but without loss of generality, in the following we will assume a linear bias model:
\begin{equation}
\label{eq:BIAS}
B(x)_k = b\, x_k\, ,
\end{equation}
where \(b\) is a constant linear bias factor. We then obtain the potential:
\begin{eqnarray}
\label{eq:POISSONIAN_POTENTIAL_B}
 \psi(\{r_k\}) &=& \frac{1}{2} ln(2\pi det(Q)) \nonumber \\
& + & \frac{1}{2}\sum_{ij} \left(r_i+\mu_i\right) Q^{-1}_{ij} \left(r_j+\mu_j\right) \nonumber \\
& - & \sum_k \left [ ln\left(\frac{(R_k\bar{N})^{N^{g}_k}}{{N^{g}_k}!}\right) + N^{g}_k\,ln(1+b\,(e^{r}-1)) \right . \nonumber \\
& &  \left .- R_k\bar{N}(1+b\,(e^{r}-1)) \right ]\, , \nonumber \\
\end{eqnarray}
and the corresponding gradient reads:
\begin{eqnarray}
\label{eq:LOGPOISSONIAN_FORCES_SIMPLE}
\frac{\partial  \psi(\{r_k\})}{\partial r_l}= \sum_{j} Q^{-1}_{lj} \left(r_j+\mu_j\right) - \left( \frac{N^{g}_l}{(1+b\,(e^{r}-1))} - R_l\bar{N}\right) b \, e^{r_l} \, . \nonumber \\ 
\end{eqnarray}
Inserting these results in equations (\ref{eq:HAMILTON1}) and (\ref{eq:HAMILTON2}) then yields the equations of motion: 
\begin{equation}
\label{eq:HAMILTON1_a}
\frac{dr_i}{dt} = \sum_j M_{ij}^{-1} \,p_j\, ,
\end{equation}
and
\begin{equation}
\label{eq:HAMILTON2_a}
\frac{dp_i}{dt} = - \sum_{j} Q^{-1}_{ij} \left(r_j+\mu_j\right) - \left( \frac{N^{g}_i}{(1+b\,(e^{r_i}-1))} - R_i\bar{N}\right) b \, e^{r_i} \, . 
\end{equation}
New points on the lognormal Poissonian posterior surface can then easily be obtained by solving for the trajectory governed by the dynamical equations (\ref{eq:HAMILTON1_a}) and (\ref{eq:HAMILTON2_a}).

%
%
%
%
%

\begin{figure*}
\centering{\resizebox{1.\hsize}{!}{\includegraphics{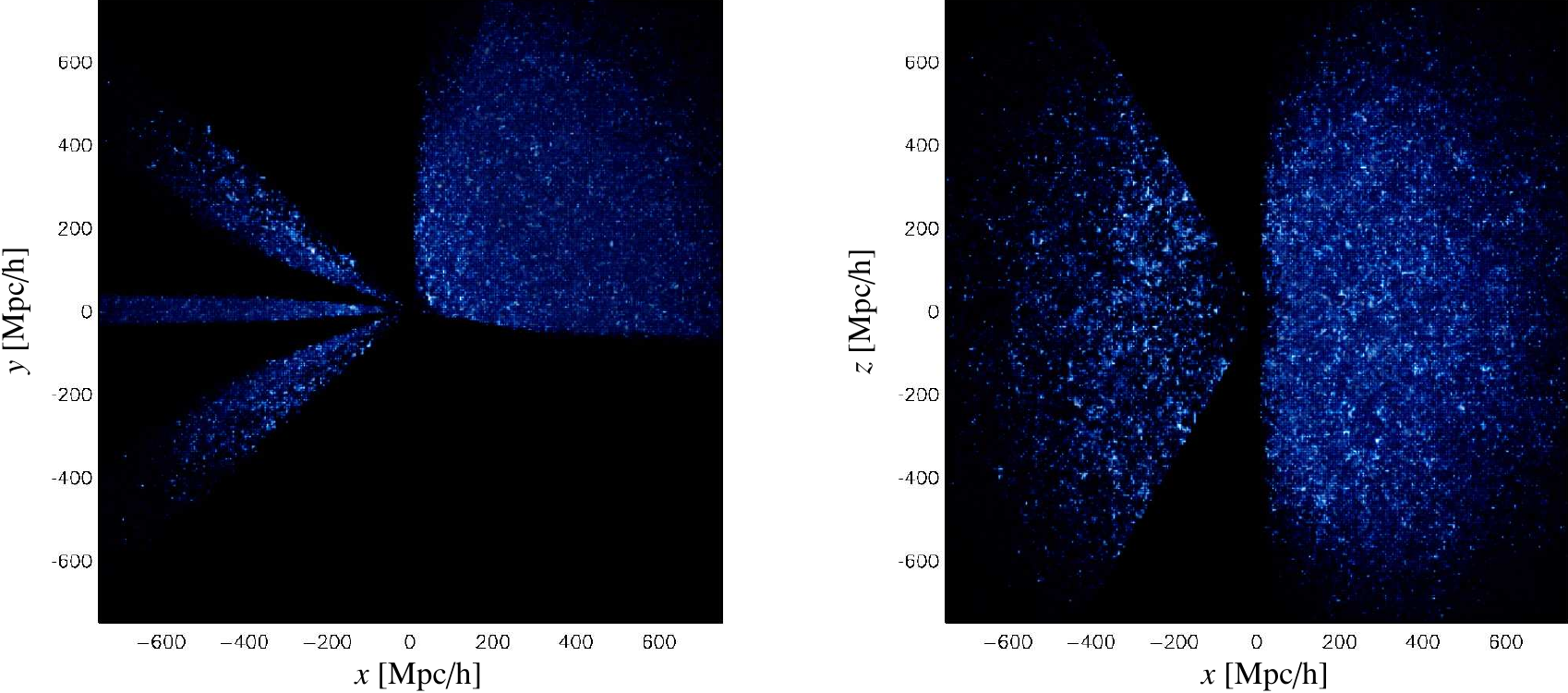}}}
\caption{Volume rendering of artificial galaxy counts, generated as described in section \ref{MOCK_Observation}. The two pannels show different projections. Effects of survey geometry and selection function are clearly visible. The observer is centered at \((0,0,0)\).}
	\label{fig:GALAXY_COUNTS}
\end{figure*}

\section{Numerical Implementation}
\label{Numerical_implementation}
Our numerical implementation of the lognormal Poissonian Sampler is named HADES (Hamiltonian Density Estimation and Sampling). It utilizes the FFTW3 library for Fast Fourier Transforms and the GNU scientific library (gsl) for random number generation \citep{FFTW05,GSL}.
In particular, we use the Mersenne Twister MT19937, with 32-bit word length, as provided by the gsl\_rng\_mt19937 routine, which was designed for Monte Carlo simulations \citep{MERSENNE_TWISTER}.

\subsection{The leapfrog scheme}
\label{LEAPFROG}
As described above, a new sample can be obtained by calculating a point on the trajectory governed by equations (\ref{eq:HAMILTON1_a}) and (\ref{eq:HAMILTON2_a}). This means that if we are able to integrate the Hamiltonian system exactly energy will be conserved along such a trajectory, yielding a high probability of acceptance.
However, the method is more general due to the Metropolis acceptance criterion given in equation (\ref{eq:acceptance_rule}). In fact, it is allowed to follow any trajectory to generate a new sample. This would enable us to use approximate Hamiltonians, which may be evaluated computationally more efficiently. Note, however, that only trajectories that approximately conserve the Hamiltonian given in equation (\ref{eq:Hamiltonian}) will result in high acceptance rates.

In order to achieve an optimal acceptance rate, we seek to solve the equations of motion exactly. For this reason, we employ a leapfrog scheme for the numerical integration. Since the leapfrog is a symplectic integrator, it is exactly reversible, a property required to ensure the chain satisfies detailed balance \citep{DUANE1987}. It is also numerically robust, and allows for simple propagation of errors. Here, we will implement the Kick-Drift-Kick scheme.
The equations of motions are integrated by making \(n\) steps with a finite stepsize \(\epsilon\), such that \(\tau=n  \epsilon\):

\begin{equation}
\label{eq:LEAPFROG1}
p_i\left(t+\frac{\epsilon}{2}\right) = p_i(t) -\frac{\epsilon}{2} \left .\frac{\partial  \psi(\{r_k\})}{\partial r_l} \right |_{r_i(t)} \, , 
\end{equation}

\begin{equation}
\label{eq:LEAPFROG2}
r_i\left(t+\epsilon \right) = r_i(t) -\frac{\epsilon}{m_i}\,p_i\left(t+\frac{\epsilon}{2}\right)  \, , 
\end{equation}

\begin{equation}
\label{eq:LEAPFROG3}
p_i\left(t+\epsilon\right) = p_i\left(t+\frac{\epsilon}{2}\right) -\frac{\epsilon}{2} \left .\frac{\partial  \psi(\{r_k\})}{\partial r_l} \right |_{r_i\left(t+\epsilon \right)} \, .  
\end{equation}
We iterate these equations until \(t=\tau\).
Also note, that it is important to vary the pseudo time interval \(\tau\), to avoid resonant trajectories. We do so by drawing \(n\) and \(\epsilon\) randomly from a uniform distribution.
For the time being we will employ the simple leapfrog scheme. However, it is possible to use higher order integration schemes, provided that exact reversibility is maintained.

\subsection{Hamiltonian mass}
\label{HMC_MASS}
The Hamiltonian sampler has a large number of adjustable parameters, namely the Hamiltonian 'mass matrix, \(M\), which  can greately influence the sampling efficiency. If the individual \(r_k\) were Gaussian distributed, a good choice for HMC masses would be to set them inversely proportional to the variance of that specific \(r_k\) \citep{TAYLOR2008}.
However, for non-Gaussian distributions, such as the lognormal Poissonian posterior, it is reasonable to use some measure of the width of the distribution \citep{TAYLOR2008}. \citet{NEAL1996} proposes to use the curvature at the peak.

In our case, we expanded the Hamiltonian given in equation (\ref{eq:POISSONIAN_POTENTIAL_B}) in a Taylor series up to quadratic order for \(|r_i| << 1\). This Taylor expansion yields a Gaussian approximation of the lognormal Poissonian posterior. Given this approximation and according to the discussion in Appendix \ref{HAMILTONIAN_MASS}, the Hamiltonian mass should be set as:
\begin{equation}
\label{eqn:HMC_MASS}
M_{ij}= Q^{-1}_{ij}-\left( \left(N^{g}_i-R_i\bar{N}\right)b - N^{g}_i b^2 \right)\delta^K_{ij}\, .
\end{equation}
However, calculation of the leapfrog scheme requires inversions of \(M\). Considering the high dimensionality of the problem, inverting and storing \(M^{-1}\) is computationally impractical. For this reason we construct a diagonal 'mass matrix' from equation (\ref{eqn:HMC_MASS}).
We found, that choosing the diagonal of \(M\), as given in equation (\ref{eqn:HMC_MASS}), in its Fourier basis yields faster convergence for the sampler than a real space representation, since it accounts for the correlation structure of the underlying density field.

\subsection{Parallelization}
\label{Parallelization}
For any three dimensional sampling method, such as the lognormal Poisson sampler or the Gibbs sampler presented in \citet{JASCHE2009}, CPU time is the main limiting factor. For this reason parallelization of the code is a crucial issue.
Since our method is a true Monte Carlo method, there exist in principle two different approaches to parallelize our code.

The numerically most demanding step in the sampling chain is the leapfrog integration with the evaluation of the potential. One could therefore parallelize the leapfrog integration scheme, which requires parallelizing the fast Fourier transform. The FFTW3 library provides parallelized fast Fourier transform procedures, and implementation of those is straightforward \citep{FFTW05}.
However, optimal speed up cannot be achieved.
The other approach relies on the fact that our method is a true Monte Carlo process, and each CPU can therefore calculate its own Markov chain.
In this fashion, we gain optimal speed up and the possibility to initialize each chain with different initial guesses. 

The major difference between these two parallelization approaches is, that with the first method one tries to calculate a rather long sampling chain, while the latter one produces many shorter chains.

\begin{figure*}
\centering{\resizebox{1.\hsize}{!}{\includegraphics{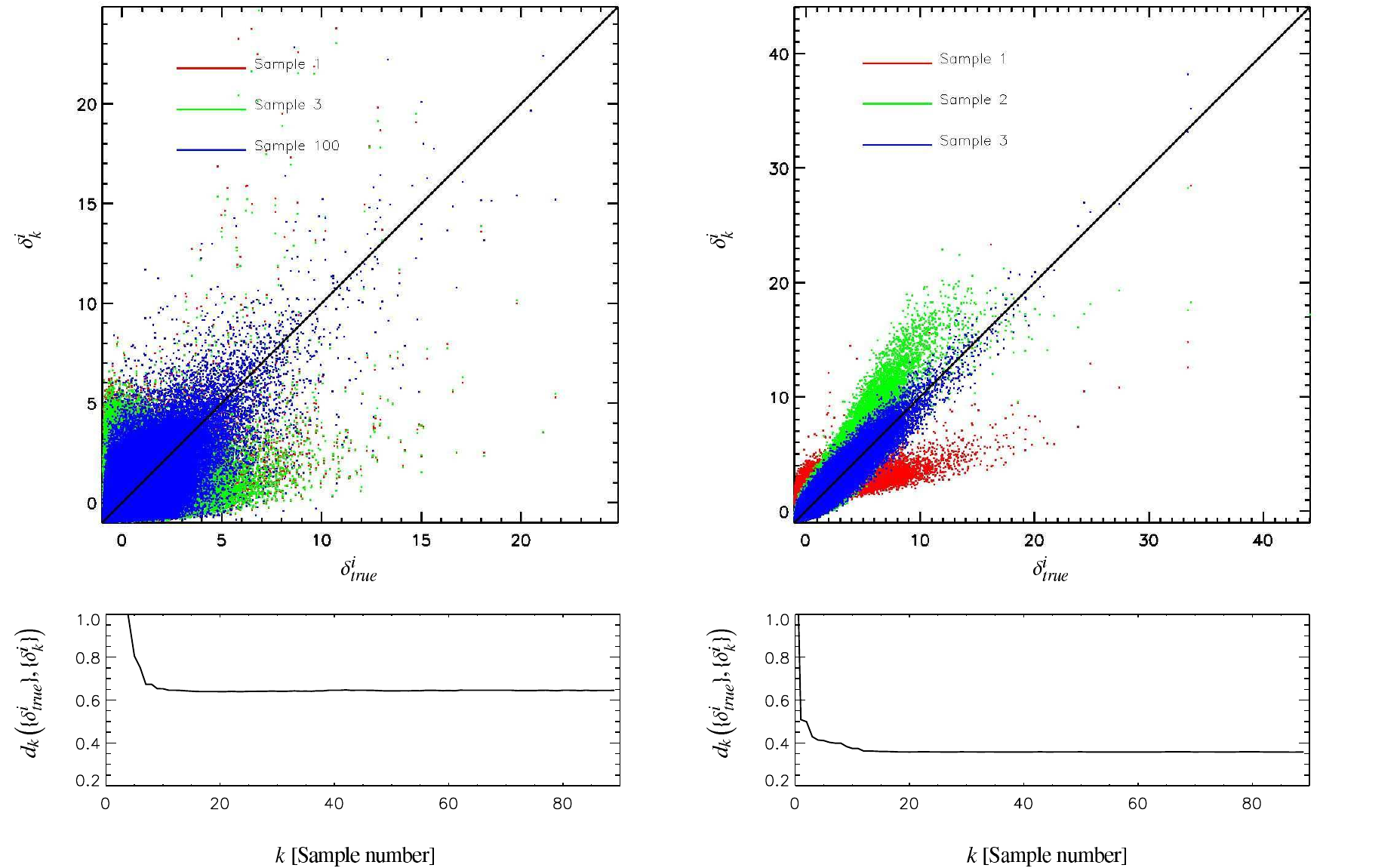}}}
\caption{Tests of the initial burn-in behavior for the two test cases of the fiducial calculation (right panels) and the full test taking into account the observational uncertainties of survey geometry and selection effects (left panels). The upper panels show successive point to point statistics between the individual samples and the true underlying mock signal. It can be seen that the successive Hamiltonian samples show increasing correlation with the true underlying signal. The lower panels show the successive Euclidean distances between samples and the true underlying signal during burn-in.}
\label{fig:BURN_IN_PERIOD_POINTSTAT}
\end{figure*}

\begin{figure*}
\centering{\resizebox{1.\hsize}{!}{\includegraphics{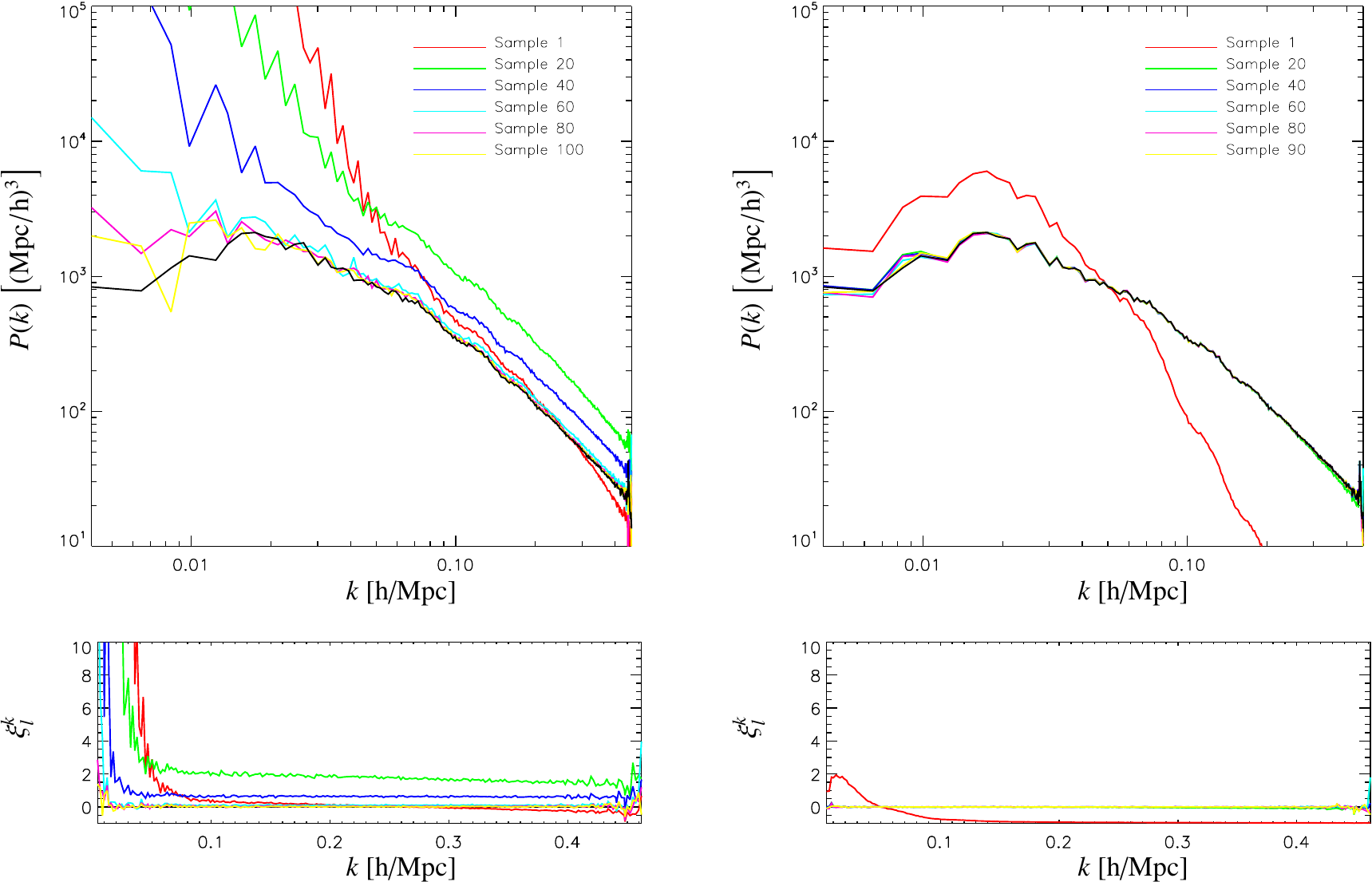}}}
\caption{Successive power spectra measured from the Hamiltonian samples during burn-in. The right panels correspond to the fiducial calculation, while the left panels display the burn-in behavior of the complete observational problem. The upper panels show the convergence of the individual sample spectra towards the spectrum of the true underlying matter field realization (black curve). The lower panels display the deviation from the true underlying spectrum \(\xi_l^k\), demonstrating good convergence at the end of the burn-in period.}
\label{fig:BURN_IN_SPECS}
\end{figure*}

\section{Testing HADES}
\label{HADES_testing}
In this section, we apply HADES to simulated mock observations, where the underlying matter signal is perfectly known. With these tests we will be able to demonstrate that the code produces results consistent with the theoretical expectation. Further more, we wish to gain insight into how the code performs in real world applications, when CPU time is limited.

\subsection{Setting up Mock observations}
\label{MOCK_Observation}
In this section we will describe a similar testing setup as described in \citet{JASCHE2009}.
For the purpose of this paper we generate lognormal random fields according to the probability distribution given in equation (\ref{eq:LogNormal_prior}). These lognormal fields are generated based on cosmological power-spectra for the density contrast \(\delta\). We generate these power spectra, with baryonic wiggles, following the prescription described in \citet{1998ApJ...496..605E} and \citet{1999ApJ...511....5E} and assuming a standard \(\Lambda\)CDM cosmology with the set of cosmological parameters (\(\Omega_m=0.24\), \(\Omega_{\Lambda}=0.76\), \(\Omega_{b}=0.04\), \(h=0.73\), \(\sigma_8=0.74\), \(n_s=1\) ).
Given these generated density fields we draw random Poissonian samples according to the Poissonian process described in equation (\ref{eq:Poissonian}).

The survey properties are described by the galaxy selection function \(F_i\) and the observation Mask \(M_i\) where the product:
\begin{equation}
\label{eq:RESPONSE_OPERATOR}
R_i=F_i\,M_i
\end{equation} 
yields the linear response operator.

The selection function is given by:
\begin{equation}
\label{eq:MOCK_GAL_SELFUNC}
F_i = \left(\frac{r_i}{r_0}\right)^b\,\left(\frac{b}{\gamma}\right)^{-b/\gamma}\,e^{\frac{b}{\gamma}-\left(\frac{r_i}{r_0}\right)^{\gamma}} \, ,
\end{equation} 
where \(r_i\) is the comoving distance from the observer to the center of the \(i\)th voxel. For our simulation we chose parameters \(b=0.6\), \(r_0=500{\unit{Mpc}}\) and \(\gamma=2\).

In figure \ref{fig:TEST_SEL_WIN} we show the selection function together with the sky mask, which defines the observed regions in the sky. The two dimensional sky mask is given in sky coordinates of right ascension and declination. We designed the observational mask to represent some of the prominent features of the Sloan Digital Sky Survey (SDSS) mask \citep[see][for a description of the SDSS data release 7]{SDSS7}.
The projection of this mask into the three dimensional volume yields the three dimensional mask \(M_i\).

Two different projections of this generated mock galaxy survey are presented in figure \ref{fig:GALAXY_COUNTS} to give a visual impression of the artificial galaxy observation.

\subsection{Burn in behavior}
\label{BURNIN}
The theory described above demonstrates that the Hamiltonian sampler will provide samples from the correct probability distribution function as long as the initial conditions for the leapfrog integration are part of the posterior surface. However, in practice the sampler is not initialized with a point on the posterior surface, and therefore an initial burn-in phase is required until a point on the correct posterior surface is identified. As there exists no theoretical criterion, which tells us when the initial burn-in period is completed, we have to test this initial sampling phase through experiments. These experiments are of practical relevance for realworld applications, as they allow us to estimate how many samples are required before the sampler starts sampling from the correct posterior distribution.
To gain intuition we set up a simple experiment, in which we set the initial guess for the lognormal field constant to unity (\(r^0_k=1\)).
Therefore, the initial samples in the chain will be required to recover structures contained in the observation. In order to gain intuition for the behavior of our nonlinear Hamiltonian sampler, we compare two cases. The first case consists of an artificial observation including selection effects and observational mask generated as described above. The second case is a comparison calculation, where we set the observation response operator \(R_i=1\). In this latter fiducial case, only shot noise remains as observational uncertainty.
It is important to note, that the individual Markov samples are unbiased in the sense that they possess the correct power information. Unlike a filter, which suppresses power in the low signal to noise regions, the Hamiltonian sampler draws true samples from the lognormal Poissonian posterior, given in equation (\ref{eq:LOGNORMALPOISSONIAN_POSTERIOR}), once burn-in is completed. Therefore, a lognormal Poissonian sample has to be understood as consisting of a true signal part, which can be extracted from the observation and a fluctuating component, which restores power lost due to the observation. This interpretation is similar to the behavior of the Gibbs sampler, as discussed in \citet{JASCHE2009}, with the exception that there is no obvious way to separate the true signal part from the variance contribution for the nonlinear Hamiltonian sampler. Hence, the lower the signal to noise ratio of the data, the higher will be the fluctuating component.

\begin{figure*}
\centering{\resizebox{1.\hsize}{!}{\includegraphics{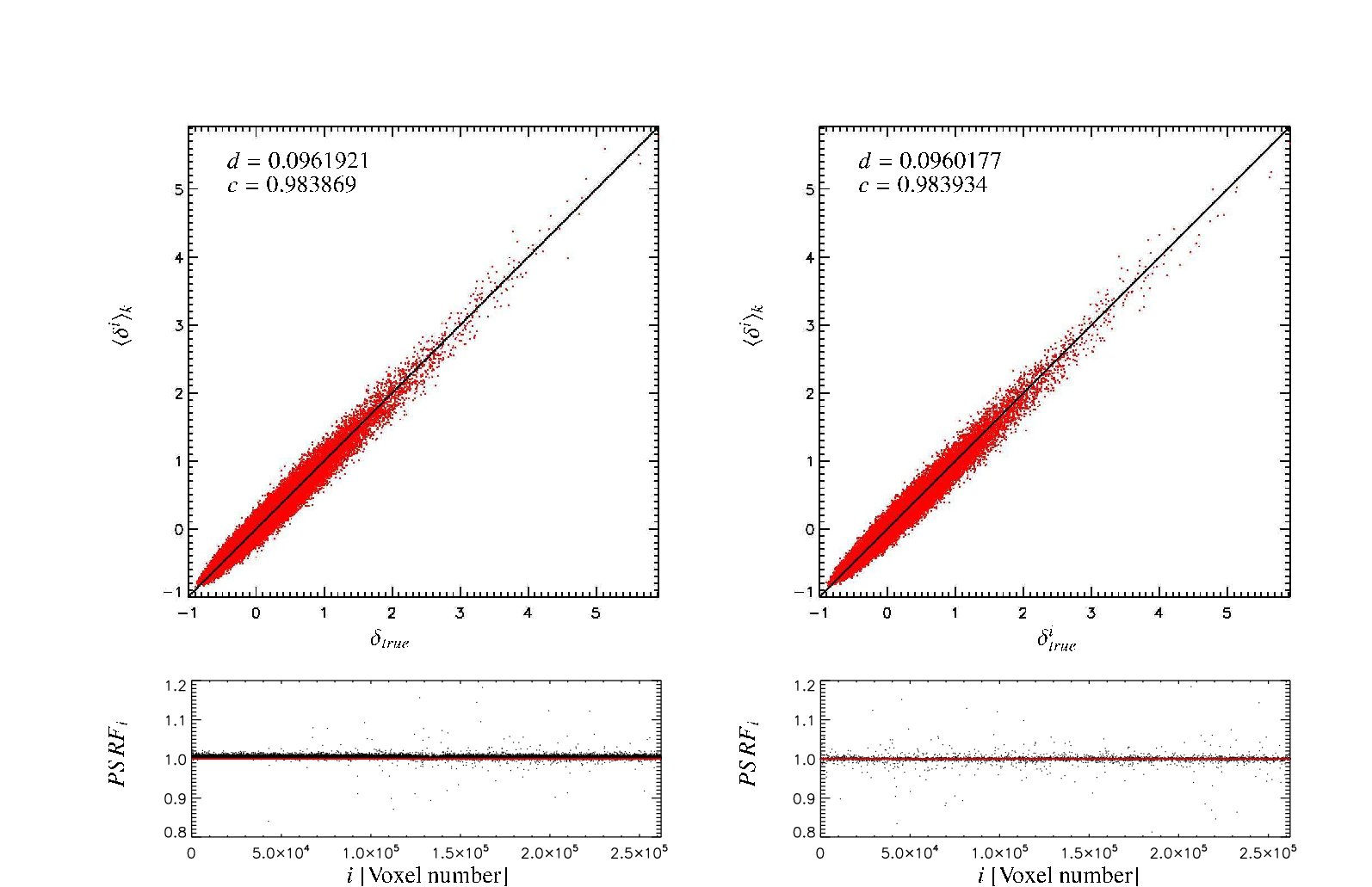}}}
\caption{The upper panels show the point to point statistic of the ensemble mean field to the true underlying density field in the observed region for the fiducial calculation (right panel) and the full observational problem (left panel). The numbers in the upper left part of the plots correspond to the Euclidean distance \(d\) and the correlation factor \(c\). In the lower panels we plotted the results of the Gelman\&Rubin convergence diagnostic for the according tests. The PSRF indicate good convergence.}
	\label{fig:GELMAN_RUBIN}
\end{figure*}

This effect can be observed in figure \ref{fig:BURN_IN_PERIOD_POINTSTAT} where we compare three successive Markov samples to the true mock signal via a point to point statistics. It can be nicely seen, that the correlation with the true underlying mock signal improves as burn-in progresses. As expected, the fiducial calculation, shown in the right panels of \ref{fig:BURN_IN_PERIOD_POINTSTAT}, has a much better correlation with the underlying true mock signal than the full observation. This is clearly owing to the fact, that the full observation introduces much more variance than in the fiducial case.
To visualize this fact further, we calculate the Euclidean distance between Markov samples and the true mock signal:
\begin{equation}
\label{eq:D_EUCLID}
d_k\left (\{\delta^i_{true}\},\{\delta^i_k\} \right) = \sqrt{\frac{1}{N} \sum_{i=1}^N \left(\delta^i_{true} - \delta^i_k \right)^2}\, ,
\end{equation} 
over the progression of the Markov chain. In the lower panels of figure \ref{fig:BURN_IN_PERIOD_POINTSTAT}, it can be observed that the Eucledian distance drops initially and then saturates at a constant minimal \(d_k\). This minimal \(d_k\) is related to the intrinsic variance contribution in the individual samples. While the variance is lower for the fiducial observation, it is higher for the full observation.

As HADES produces unbiased samples, we can gain more detailed insight into the initial burn-in phase of the Markov chain, by following the evolution of successive power-spectra measured from the samples. In addition, we measure the deviation \(\xi_l^k\) of the sample power spectra \(P_l^k\) to the power spectrum of the true mock matter field realization \(P_l^{true}\) via:
\begin{equation}
\label{eq:BURN_IN}
\xi_l^k = \frac{P_l^k-P_l^{true}}{P_l^{true}}\, .
\end{equation}
Figure \ref{fig:BURN_IN_SPECS} demonstrates that HADES completes burn-in after \(\sim 20\) samples in the case of the fiducial calculation (right panels). However, the burn-in history for the full observation (left panels) reveals a more interesting behavior.

Initially, the power spectra show huge excursions at large scales. This is due to the observational mask and the fact, that initially these regions are dominated by the constant initial guess (\(r^0_k=1\)). It is interesting to note, that the first sample seems to be much closer to the true underlying power specrum at the smaller scales, while the 20th samples is much further away. This clearly demonstrates the nonlinear behavior of the lognormal Poissonian sampler. We observe, that with iterative correction of the large scale power, the entire power spectrum progressively approaches the true mock power spectrum. This can be seen nicely in the lower left panel of figure \ref{fig:BURN_IN_SPECS}. 
After one hundred samples have been calculated the true mock power spectrum is recovered for all following samples. Thus, the initial burn-in period for a realistic observational setting can be expected to be on the order of \(100\) samples. Such a burn-in period is numerically not very demanding, and can easily be achieved in even higher resolution calculations.

Further, we ran a full Markov analysis for both test cases, by calculating \(20000\) samples with a resolution of \(64^3\) voxels. We then estimate the ensemble mean and compared the recovered density field in the observed region via a point to point statistic to the true underlying mock signal. The results are presented in the upper panels of figure \ref{fig:GELMAN_RUBIN}. It can be seen that both results are strongly correlated with the true underlying signal. To emphasize this fact, we also calculate the correlation factor given as:
\begin{equation}
\label{eq:correlation_factor}
c = \frac{ \sum_{i=0}^{N-1} \delta^{true}_i\, \delta^{mean}_i} {\sqrt{ \sum_{i=0}^{N-1} \left(\delta^{true}_i\right)^2} \sqrt{\sum_{i=0}^{N-1}\left(\delta^{mean}_i\right)^2}}\, .
\end{equation}
The correlation factors for the two test scenarios are also given in figure \ref{fig:GELMAN_RUBIN}. They clearly demonstrate, that the Hamiltonian sampler was able to recover the underlying density field to high accuracy in both cases.

\subsection{Convergence}
\label{Convergence}
Testing the convergence of Markov chains is subject of many discussions in literature \citep[see e.g.][]{HEIDELBERGER1981,GELMAN1992,GEWEKE1992,RAFTERY1995,COWLES1996,HANSON2001,DUNKLEY2005}. In principle, there exist two categories of possible diagnostics. The methods of the first category rely on comparing inter chain quantities between several chains while others try to estimate the convergence behavior from inter chain quantities within a single chain.
In this paper we use the widely used Gelman\&Rubin diagnostic, which is based on multiple simulated chains by comparing the variances within each chain and the variance between chains \citep{GELMAN1992}. In particular, we calculate the potential scale reduction factor (PSRF) (see Appendix \ref{Gelman_Rubin_diagnostic} for details). A large PSRF indicates that the inter chain variance is substantially greater than the intra chain variance and longer chains are required. Once the PSRF approaches unity, one can conclude that each chain has reached the target distribution.

We calculated the PSRF for each voxel of our test cases for chains with length \(N_{samp}=20000\). The results for the two tests, as discussed above, are presented in figure \ref{fig:GELMAN_RUBIN}. They clearly indicate the convergence of the Markov chains.

For the time being we use the Gelman\&Rubin statistic to test convergence because of technical simplicity, although for the expense of having to calculate at least two chains. In the future we plan to explore other convergence diagnostics. In particular we are aiming at including intra chain methods as proposed in \citet{HANSON2001} or \citet{DUNKLEY2005}. This would allow us to detect convergence behavior within the chain during burn-in. Such a convergence criterion could then be used to adjust the Hamiltonian masses for optimal sampling efficiency, as was proposed in \citet{TAYLOR2008}.

\begin{figure*}
\centering{\resizebox{1.\hsize}{!}{\includegraphics{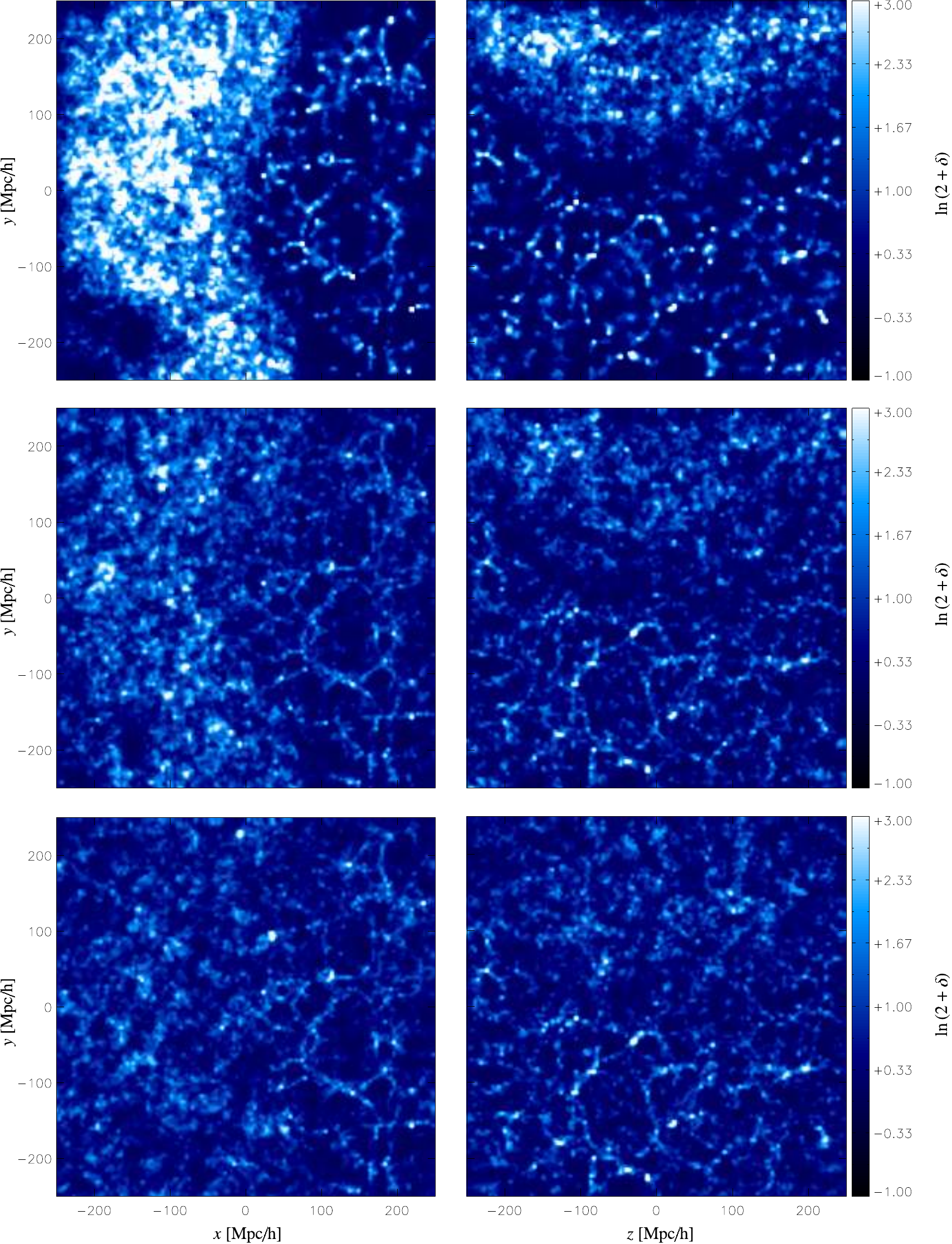}}}
\caption{Slices through density samples during the initial burn-in phase. The upper panels correspond to the first sample, middle panels show the tenth sample and the lower panels present the hundredth sample. Left and right panels show two different slices through the corresponding sample. It can be seen that during the initial burn-in phase power equalizes between the observed and unobserved regions. Successive samples recover finer and finer details.}
	\label{fig:BURNIN_CROTON}
\end{figure*}

\subsection{Testing with simulated galaxy surveys}
\label{Discussion}
In this section, we describe the application of HADES to a mock galaxy survey based on the millennium run \citep{CROTON2006}. The intention of this exercise is to test HADES in a realistic observational scenario. In particular, we want to demonstrate that HADES is able to reconstruct the fully evolved nonlinear density field of the N-body simulation. 
The mock galaxy survey consists of a set of comoving galaxy positions distributed in a 500 Mpc box. To introduce survey geometry and selection effects, we virtually observe these galaxies through the sky mask and according to the selection function described in section \ref{MOCK_Observation}. The resulting galaxy distribution is then sampled to a \(128^3\) grid. This mock observation is then processed by HADES, which generates \(20000\) lognormal Poissonian samples.

In figure \ref{fig:BURNIN_CROTON} we present successive slices through density samples of the initial burn-in period. As can be seen, the first Hamiltonian sample (upper panels in figure \ref{fig:BURNIN_CROTON}) is largely corrupted by the false density information in the masked regions. This is due to the fact, that the Hamiltonian sampler cannot be initialized with a point on the posterior surface. The initial samples are therefore required to identify a point on the according posterior surface. As can be seen, the power in the unobserved and observed regions equalizes in the following samples.
Also note, that the first density sample depicts only very coarse structures. However, subsequent samples resolve finer and finer details.
With the hundredth sample burn-in is completed. The lower panels of figure \ref{fig:BURNIN_CROTON} demonstrate, that the Hamiltonian sampler nicely recovers the filamentary structure of the density field.

Being a fully Bayesian method, the Hamiltonian sampler does not aim at calculating only a single estimate, such as a mean or maximum a postiori value, it rather produces samples from the full lognormal Poissonian posterior. Given these samples we are able to calculate any desired statistical summary. In particular, we are able to calculate the mean and the according variance of the Hamiltonian samples.

\begin{figure*}
\centering{\resizebox{1.\hsize}{!}{\includegraphics{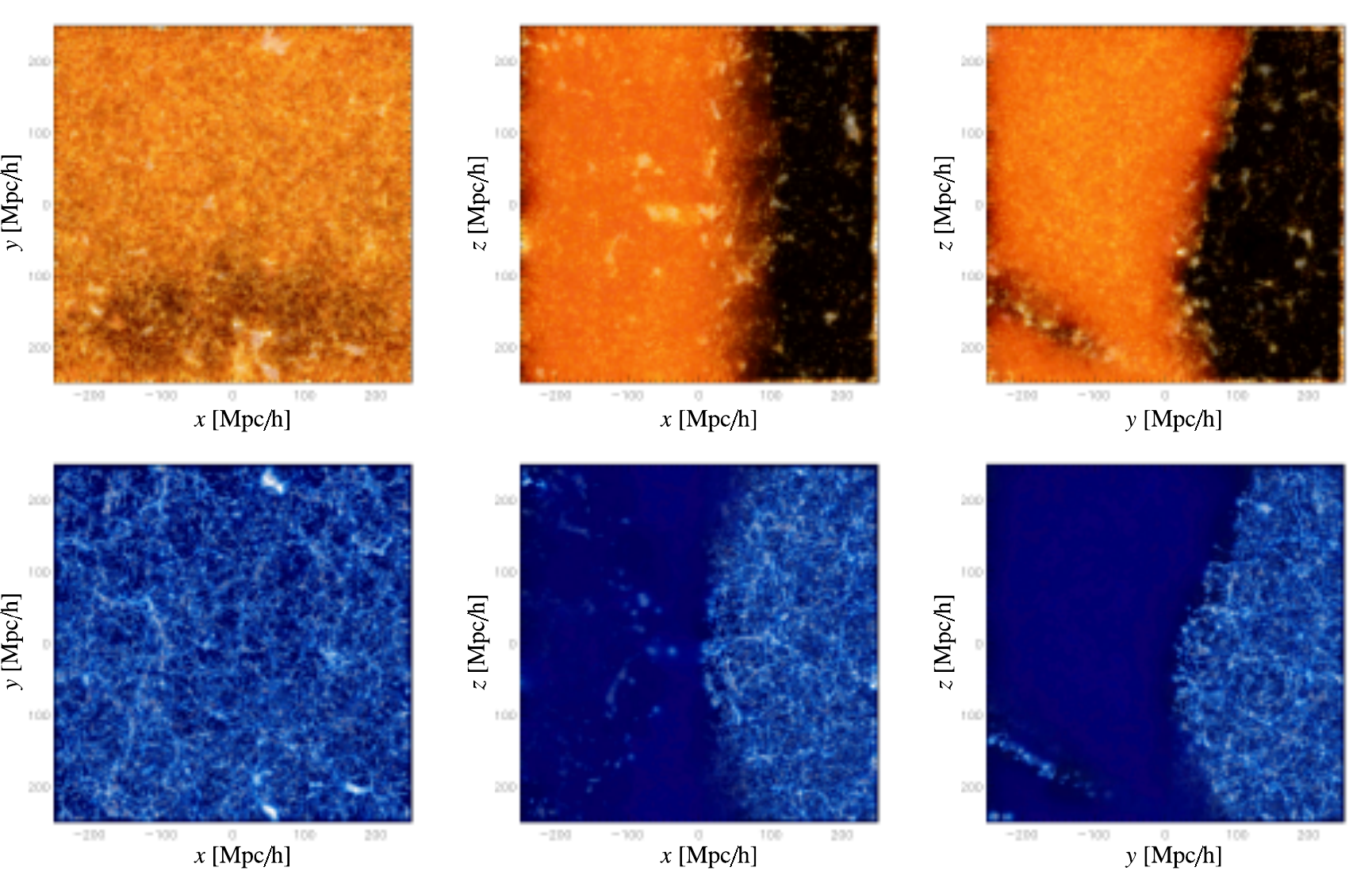}}}
	\caption{Volume rendering of the ensemble variance (upper panels) and the ensemble mean (lower panels) obtained from the mock galaxy catalog analysis for three different perspectives. The mean shows filigree structures which have been recovered. It can also be seen that the ensemble variance reflects the Poissonian behavior of the noise. High density regions in the ensemble mean field correspond to regions with high variance as is expected for a Poissonian shot noise contribution.}
	\label{fig:CROTON_MEAN_VAR}
\end{figure*}

In figure \ref{fig:CROTON_MEAN_VAR} we show three different volume renderings of the ensemble mean density and the according ensemble variance fields. It can be seen that the variance projections nicely reflect the Poissonian noise structure. Comparing high density regions in the ensemble mean projections to the corresponding positions in the variance projections, reveals a higher variance contribution for these regions, as expected for Poissonian noise. This demonstrates, that our method allows us to provide uncertainty information for any resulting final estimate.

\section{Summary and Conclusion}
\label{Discussion}
In this work we introduced the Hamiltonian Monte Carlo sampler for nonlinear large scale structure inference and demonstrated its performance in a variety of tests. 
As already described above, according to observational evidence and theoretical considerations, the posterior for nonlinear density field inference is adequately represented by a lognormal Poissonian distribution, up to overdensities of \(\delta \sim 100\). Hence, any method aiming at precision estimation of the fully evolved large scale structure in the Universe needs to handle the nonlinear relation between observations and the signal we seek to recover. 
The Hamiltonian Monte Carlo sampler, presented in this work, is a fully Bayesian method, and as such tries to evaluate the lognormal Poissonian posterior, given in equation \ref{eq:LOGNORMALPOISSONIAN_POSTERIOR}, via sampling. In this fashion, the scientific output of the method is not a single estimate, but a sampled representation of the multidimensional posterior distribution. Given this representation of the posterior any desired statistical summary, such as mean, mode or variances can easily be calculated. Further, any uncertainty can seamlessly be propagated to the finally estimated quantities, by simply applying the according estimation procedure to all Hamiltonian samples.

Unlike conventional Metropolis Hastings algorithms, which move through the parameter space by random walk, the Hamiltonian Monte Carlo sampler suppresses random walk behavior by following a persistent motion. The HMC exploits techniques developed to follow classical dynamical particle motion in potentials, which, in the absence of numerical errors, yield an acceptance probability of unity.
Although, in this work we focused on the use of the lognormal Poissonian posterior, the method is more general. The discussion of the Hamiltonian sampler in section \ref{HAMILTONIAN_SAMPLING}, demonstrates that the method can in principle take into account a broad class of posterior distributions.

In section \ref{HADES_testing}, we demonstrated applications of the method to mock test cases, taking into account observational uncertainties such as selection effects, survey geometries and noise. These tests were designed to study the performance of the method in real world applications.

In particular, it was of interest to establish intuition for the behavior of the Hamiltonian sampler during the initial burn-in phase. Especially, the required amount of samples before the sampler starts drawing samples from the correct posterior distribution was of practical relevance. The tests demonstrated, that for a realistic setup, the initial burn-in period is on the order of \(\sim 100\) samples.

Further, the tests demonstrated that the Hamiltonian sampler produces unbiased samples, in the sense that each sample possesses correct power. Unlike a filter, which suppresses the signal in low signal to noise regions, the Hamiltonian sampler nonlinearly augments the poorly or not observed regions with correct statistical information. In this fashion, each sample represents a complete matter field realization consistent with the observations. 

The convergence of the Markov chain was tested via a Gelman\&Rubin diagnostic. We compared the intra and inter chain variances of two Markov chains each of length \(20000\) samples. The estimated PSRF indicated good convergence of the chain. This result demonstrates, that it is possible to efficiently sample from non-Gaussian distributions in very high dimensional spaces.

In a final test the method was applied to a realistic galaxy mock observation based on the millennium run \citep{CROTON2006}. Here we introduced again survey geometry and selection effects and generated \(20000\) samples of the lognormal Poissonian posterior. The results nicely demonstrate that the Hamiltonian sampler recovers the filamentary structure of the underlying matter field realization. For this test we also calculated the ensemble mean and the corresponding ensemble variance of the Hamiltonian samples, demonstrating that the Hamiltonian sampler also provides error information for a final estimate.

To conclude, in this paper we present a new and numerically efficient Bayesian method for large scale structure inference, and its numerical implementation HADES. HADES provides a sampled representation of the very high dimensional non-Gaussian large scale structure posterior, conditional on galaxy observations. This permits us to easily calculate any desired statistical summary, such as mean, mode and variance. In this fashion HADES is able to provide uncertainty information to any final quantity estimated from the Hamiltonian samples.
The method, as presented here, is very flexible and can easily be extended to take into account additional nonlinear observational constraints and joint uncertainties.

In summary, HADES, in its present form, provides the basis for future nonlinear high precision large scale structure analysis.

\section*{Acknowledgments}
We would like to thank Rainer Moll and Bj\"{o}rn Malte Sch\"{a}fer for usefull discussions and support with many valuable numerical gadgets. Further, we thank R. Benton Metcalf, for useful remarks and commentaries, and Torsten A. En\ss lin for encouraging us to persue this project. Special thanks also to Mar\'{i}a \'{A}ngeles Bazarra-Castro for helpful assistance during the course of this project.
We also thank the "Transregional Collaborative Research Centre TRR 33 - The Dark Universe" for the support of this work.

\bibliography{paper}
\bibliographystyle{mn2e}

\appendix
\section{Hamiltonian Masses}
\label{HAMILTONIAN_MASS}
The Hamiltonian sampler can be extremely sensitive to the choice of masses. To estimate a good guess of Hamiltonian masses we follow a similar approach as suggested in \citet{TAYLOR2008}. According to the leapfrog scheme, given in equations (\ref{eq:LEAPFROG1}), (\ref{eq:LEAPFROG2}) and (\ref{eq:LEAPFROG3}), a single application of the leapfrog method can be written in the form:
\begin{equation}
\label{eqn:LEAPFROG_STEP_1_full}
p_i(t+\epsilon)=p_i(t) -\frac{\epsilon}{2} \left ( \left .\frac{\partial \Psi(r)}{\partial r_i}\right|_{r(t)} + \left .\frac{\partial \Psi(r)}{\partial r_i}\right|_{r(t+\epsilon)} \right)
\end{equation}

\begin{equation}
\label{eqn:LEAPFROG_STEP_2_full}
r_i(t+\epsilon)=r_i(t) +\epsilon \sum_j M^{-1}_{ij}\, p_j(t) -\frac{\epsilon^2}{2} \sum_j M^{-1}_{ij} \left .\frac{\partial \Psi(r)}{\partial r_j}\right|_{r(t)} \, . 
\end{equation}
We will then approximate the forces given in equation (\ref{eq:LOGPOISSONIAN_FORCES_SIMPLE}) for \(r_i << 1\):
\begin{eqnarray}
\label{eq:APPROX_FORCES}
\frac{\partial \psi(r)}{\partial r_l} &=& \sum_{j} Q^{-1}_{lj} \left(r_j+\mu_j\right) - \left( \frac{N^{g}_l}{(1+b\,(e^{r}-1))} - R_l\bar{N}\right) b \, e^{r_l}  \nonumber \\ 
&\approx&\sum_{j} Q^{-1}_{lj} \left(r_j+\mu_j\right) - \left[ \left(N^{g}_l-R_l\bar{N}\right)b \right . \nonumber \\
& & \left .+\left( \left(N^{g}_l-R_l\bar{N}\right)b - N^{g}_l b^2 \right)\, r_l \right] \, . \nonumber \\ 
&=& \sum_{j} \left [Q^{-1}_{lj}-\left( \left(N^{g}_l-R_l\bar{N}\right)b - N^{g}_l b^2 \right)\delta^K_{lj} \right ] r_j \nonumber \\
& & + \sum_{j} Q^{-1}_{lj} \mu_j - \left(N^{g}_l-R_l\bar{N}\right)b \, .\nonumber \\
\end{eqnarray}
By introducing:
\begin{equation}
\label{eqn:A_MATRIX}
A_{lj}= Q^{-1}_{lj}-\left( \left(N^{g}_l-R_l\bar{N}\right)b - N^{g}_l b^2 \right)\delta^K_{lj}
\end{equation}
and
\begin{equation}
\label{eqn:D_VECTOR}
D_{l}= \sum_{j} Q^{-1}_{lj} \mu_j - \left(N^{g}_l-R_l\bar{N}\right)b \, ,
\end{equation}
equation (\ref{eq:APPROX_FORCES}) simplifies to:
\begin{eqnarray}
\label{eq:APPROX_FORCES_A}
\frac{\partial \psi(r)}{\partial r_l} &=& \sum_{j} A_{lj}\,r_j+D_l \, .
\end{eqnarray}
Introducing this approximation into equations (\ref{eqn:LEAPFROG_STEP_1_full}) and (\ref{eqn:LEAPFROG_STEP_2_full}) yields: 

\begin{eqnarray}
\label{eqn:LEAPFROG_STEP_1_full_a}
p_i(t+\epsilon)&=&\sum_{m} \left[\delta^{K}_{im}-\frac{\epsilon^2}{2} \sum_j A_{ij} M^{-1}_{jm}\right]\, p_m(t) \nonumber \\
& &-\epsilon\sum_{j} A_{ij} \sum_p\left[\delta^{K}_{jp}-\frac{\epsilon^2}{4}\sum_m M^{-1}_{jm}\,A_{mp}\right]\,r_p(t)\nonumber \\
& & - \frac{\epsilon}{2} \sum_m \left [ \delta^{K}_{im} - \frac{\epsilon^2}{2} \sum_j A_{ij} M^{-1}_{jm} \right ] D_m\nonumber \\
\end{eqnarray}
and
\begin{eqnarray}
\label{eqn:LEAPFROG_STEP_2_full_a}
r_i(t+\epsilon)&=&\epsilon \sum_j M^{-1}_{ij} \, p_j(t) \nonumber \\
& &+\sum_{m}\left(\delta^K_{im} -\frac{\epsilon^2}{2}\sum_j M^{-1}_{ij} \,A_{jm}\right)r_m(t)\nonumber \\
& & -\frac{\epsilon^2}{2}\sum_j M^{-1}_{ij}D_j\, . \nonumber \\ 
\end{eqnarray}
This result can be rewritten in matrix notation as:
\begin{equation}
\label{eqn:vector}
\left (\begin{array}{c} r(t+\epsilon)\\ p(t+\epsilon)\end{array}\right ) = T \left (\begin{array}{c} r(t)\\ p(t)\end{array}\right ) - \frac{\epsilon^2 }{2}\left(\begin{array}{c} M^{-1}\,D\\ \epsilon \left [ \rm{I} -\frac{\epsilon^2}{2}A\,M^{-1}\right] D\end{array}\right )  \, ,
\end{equation}
where the matrix \(T\) is given as:
\begin{equation}
\label{eqn:vector}
T=\left (\begin{array}{cc} \left [\rm{I}-\frac{\epsilon^2}{2} M^{-1} A\right] &  \epsilon M^{-1}\\ -\epsilon\,A\left[\rm{I}-\frac{\epsilon^2}{4}M^{-1}\,A\right] & \left[\rm{I}-\frac{\epsilon^2}{2} A\,M^{-1}\right]\end{array}\right ) \, ,
\end{equation}
with \(\rm{I}\) being the identity matrix.
Successive applications of the leapfrog step yield the following propagation equation:
\begin{equation}
\label{eqn:iterative_vector}
\left (\begin{array}{c} r^n\\ p^n\end{array}\right ) = T^n \left (\begin{array}{c} r^0\\ p^0\end{array}\right ) - \frac{\epsilon^2 }{2} \left [\sum^{n-1}_{i=0} T^i \right] \left(\begin{array}{c} M^{-1}\,D\\ \epsilon \left [ \rm{I} -\frac{\epsilon^2}{2}A\,M^{-1}\right] D\end{array}\right )  \, .
\end{equation}
This equation demonstrates, that there are two criteria to be fulfilled if the method is to be stable under repeated application of the leapfrog step.
First we have to ensure, that the first term of equation (\ref{eqn:iterative_vector}) does not diverge. This can be fulfilled if the eigenvalues of \(T\) have unit modulus. The eigenvalues \(\lambda\) are found by solving the characteristic equation:
\begin{equation}
\label{eqn:characteristic_equation}
det\left[ \rm{I}\, \lambda^2 - 2\,\lambda \left ( \rm{I} -\frac{\epsilon^2}{2}A\,M^{-1}\right) + \rm{I} \right ]=0 \, .
\end{equation}
Note, that this is a similar result to what was found in \citet{TAYLOR2008}.
Our aim is to explore the parameter space rapidly, and therefore we wish to choose the largest \(\epsilon\) still compatible with the stability criterion. However, any dependence of equation (\ref{eqn:characteristic_equation}) also implies, that no single value of \(\epsilon\) will meet the requirement for every eigenvalue to have unit modulus. For this reason we choose:
\begin{equation}
\label{eqn:stability_condition}
A=M \, .
\end{equation}
We then yield the characteristic equation:
\begin{equation}
\label{eqn:characteristic_equation_a}
\left[ \lambda^2 - 2\,\lambda \left ( 1 -\frac{\epsilon^2}{2}\right) + 1 \right ]^N=0 \, ,
\end{equation}
where \(N\) is the number of voxels. This yields the eigenvalues:
\begin{equation}
\label{eqn:eigenvals}
\lambda = \pm \, i \sqrt{1-\left[1 -\frac{\epsilon^2}{2} \right]^2} +\left[1 -\frac{\epsilon^2}{2} \right] \, ,
\end{equation}
which have unit modulus for \(\epsilon \le 2 \).
The second term in equation (\ref{eqn:iterative_vector}) involves evaluation of the geometric series \(\sum^{n-1}_{i=0} T^i\).
However, the geometric series for a matrix converges if and only if \(| \lambda _i| < 1 \) for each \( \lambda _i \) eigenvalue of \(T\). 
This clarifies, that the nonlinearities in the Hamiltonian equations generally do not allow for arbitrary large pseudo time steps \(\epsilon\). In addition, for practical purposes we usually restrict the mass matrix to the diagonal of equation (\ref{eqn:A_MATRIX}).
For these two reasons, in practize, we choose the pseudo timestep \(\epsilon\) as large as possible while still obtaining a reasonable rejection rate.

\section{Gelman\&Rubin diagnostic}
\label{Gelman_Rubin_diagnostic}
The Gelman\&Rubin diagnostic is a multichain convergence test \citep{GELMAN1992}. It is based on analyzing multiple Markov chains by comparing intra chain variances, within each chain, and inter chain variances. A large deviation between these two variances indicates nonconvergence of the Markov chain.
Let \(\{\phi^k\}\), where \(k=1,...,N\), be the collection of a single Markov chain output. The parameter \(\phi^k\) is the \(k\)th sample of the Markov chain. Here, for notational simplicity, we will assume \(\phi\) to be single dimensional. To test convergence with the Gelman\&Rubin statistic, one has to calculate \(M\) parallel MCMC chains, which are initialized from different parts of the target distribution. After discarding the initial burn-in samples, each chain is of length \(n\). We can then label the outputs of various chains as \(\phi^k_m\), with \(k=1,...,n\) and \(m=1,...,M\).
The inter chain variance \(B\) can then be calculated as:
\begin{equation}
\label{eqn:INTRACHAINVARIANCE}
B = \frac{n}{M-1}\sum_{m=1}^M \left(\theta_m-\Omega\right)^2 \, ,
\end{equation}
where \(\theta_m\) is given as:
\begin{equation}
\label{eqn:INTRACHAINVARIANCE_thetam}
\theta_m = \frac{1}{n} \sum_{k=1}^n \phi^k_m \, , 
\end{equation}
and \(\Omega\) as:
\begin{equation}
\label{eqn:INTRACHAINVARIANCE_thetam}
\Omega = \frac{1}{M} \sum_{m=1}^M \theta_m \, . 
\end{equation}
Then the intra chain variance can be calculated as:
\begin{equation}
\label{eqn:INTERCHAINVARIANCE}
W = \frac{1}{M}\sum_{m=1}^M \Gamma^2_m \, ,
\end{equation}
with :
\begin{equation}
\label{eqn:INTERCHAINVARIANCE_GAMMA}
\Gamma^2_m = \frac{1}{n-1} \sum_{k=1}^n (\phi^k_m-\theta_m)^2\, . 
\end{equation}
With the above definition the marginal posterior variance can be estimated via:
\begin{equation}
\label{eqn:POSTERIOR_VAR}
V = \frac{n-1}{n}\, W + \frac{M+1}{nM}\,B\, . 
\end{equation}
If all \(M\) chains have reached the target distribution, this posterior variance estimate should be very close to the intra chain variance \(W\). For this reason, one expects the ratio \(V/W\) to be close to \(1\). The square root of this ratio is reffered to as the potential scale reduction factor (PSRF):
\begin{equation}
\label{eqn:PSRF}
PSRF = \sqrt{\frac{V}{W}}\, . 
\end{equation}
If the PSRF is close to one, one can conclude that each chain has stabilized, and has reached the target distribution  \citep{GELMAN1992}.

\bsp

\label{lastpage}

\end{document}